\documentclass{ctr_summer}

\usepackage{ctrfont}
\usepackage{natbib}
\usepackage{undertilde}
\usepackage{color}

\usepackage{graphicx}
\usepackage{epstopdf}

\usepackage{epsfig}
\usepackage{amsmath,rotating}
\usepackage{dashrule}
\usepackage{undertilde}
\usepackage{subfigure}

\usepackage{url}

\usepackage{color}

\title{Airfoil trailing-edge noise source identification using large-eddy simulation and wavelet transform}

\shorttitle{Airfoil noise using LES and wavelet transform}

\author{S. Lee\footnote[1]{Department of Mechanical and Aerospace Engineering, University of California, Davis}, D. Kang\footnotemark[1], D. Brouzet \and S. K. Lele}

\shortauthor{Lee et~al.}

\begin{document}
\pagenumbering{gobble}

\setcounter{page}{1}

\maketitle

Airfoil noise is predicted and analyzed using wall-resolved large-eddy simulations and wavelet transforms for a NACA 0012 airfoil at
a Mach number of 0.06 and a Reynolds number of 400,000 using a stair-strip forced transition and a natural transition. At a high angle
of attack, vortex shedding and a laminar separation bubble (LSB) occur
on the suction side. The LSB triggers the flow transition for both the forced
and natural transition cases. The wavelet thresholding and
denoising algorithm is used to decompose the pressure fields into
the coherent or denoised pressure and the incoherent or background
noise pressure. This denoising technique provides a clear picture
of true noise generation and propagation. It also reveals the
dominant noise source at specific frequencies when multiple noise
sources are present. In another usage, the wavelet thresholding
algorithm with down-sampling separates noise on the basis of flow structures. For example, the wavelet method separates
noise between low-frequency vortex shedding noise and
high-frequency LSB noise as well as
trailing-edge noise. Finally, the wavelet transform is used to
decompose the hydrodynamic and acoustic pressure components near
the surface using the coherence between near-field pressure and
far-field pressure. Overall, the wavelet-based decomposition is a
valuable tool to study and reveal the mechanisms of airfoil noise
generation. \hrule

\section{Introduction}
Turbulent boundary layer trailing-edge noise is an important fluid
mechanics problem involving a turbulent flow and its interaction
with a solid body \citep{Lee_2021}. This noise is generated when
turbulent boundary layer pressure fluctuations are scattered by a
sharp edge or they experience a sudden change in the boundary
conditions. This problem occurs for many aerospace and industrial
applications, such as aircraft wings, rotorcrafts, wind turbines,
or propellers. In particular, the recent advent of electric
vertical takeoff and landing or urban air mobility aircraft
further increased interest in this trailing-edge noise because
this noise becomes more apparent at low RPM with more blades
\citep{Greenwood_2022,Li_2020,Li_2021,Li_2022}.

Large-eddy simulations (LES) have been applied to predict
trailing-edge noise at low Reynolds numbers
\citep{Wolf_2012,Bodling_2019}. The accuracy of numerical
simulations depends on many parameters, including the grid
resolution, time step, and numerical scheme. For example, if the
grid resolution is not sufficient, unphysical numerical noise or
background noise occurs at high frequencies \citep{Brouzet_2020}.
It is important to identify and eliminate this numerical noise in
the time-domain pressure fields and frequency-domain spectra to
capture only true or physical noise sources.


Although trailing-edge noise is an important component of airfoil
broadband noise, especially at high frequencies, there are also
other competing noise sources, such as vortex shedding noise,
laminar separation bubble (LSB) instability noise, tip noise,
stall noise, and tripping noise \citep{Brooks_1989}. In
particular, vortex shedding noise is observed as tonal noise at low
frequencies. LSB and tripping noise that exhibit tonal and
broadband noise characteristics at high frequencies are affected
by flow conditions, airfoil geometries, angles of attack,
numerical tripping methods, and so on. It is desirable to separate the
acoustic fluctuations arising from physically different noise
sources.

Based on the acoustic analogy, the pressure fluctuation on the
surface is the dipole noise source \citep{Amiet_1976}. The
unsteady and random pressure fluctuations associated with
trailing-edge noise include two major components: hydrodynamic
turbulent pressure fluctuations and acoustic pressure fluctuations
due to the scattering by a trailing edge. In incompressible flow
simulations, the acoustic scattering is not captured, so it should
be separately accounted for. The hydrodynamic wall pressure
fluctuations can be independently determined as well
\citep{Lee_2018,Lee_2019}. For compressible flow simulations, both
the hydrodynamic and acoustic parts are captured. However, it is
difficult to distinguish these two components and analyze their
relative contributions to far-field acoustics. As new noise
reduction techniques, such as serrations or finlets, are being
developed, distinguishing these source components becomes
important, as this distinction reveals an exact source of noise
generation and provides means to control such noise.

Wavelet transform methods have proved useful to decompose the
pressure fields in turbulent flows
\citep{Farge_1992,Schneider_2010} and, more recently, in jet
aero-acoustics \citep{Mancinelli_2017}. In the wavelet thresholding
technique, the magnitudes of wavelet coefficients are divided into
two categories based on the selection of an appropriate threshold.
These categories were identified as coherent and incoherent fields
\citep{Schneider_2010}. For example, the wavelet thresholding
denoising technique was used to filter out nearly Gaussian white
noise, which was identified as the incoherent field, and to retain
only the coherent vortical fields
\citep{Azzalini_2005,Nguyen_2012}.

In our research, we use the wavelet-based decomposition method to
address various important questions in airfoil noise. The first
goal is to identify and eliminate numerical noise and provide the
denoised pressure fields. This will help reveal true noise sources
in numerical simulations. The second goal is to separate the
acoustic pressure fields depending on the flow structures or noise
sources. Finally, we separate the hydrodynamic pressure from
the acoustic pressure in airfoil noise using the wavelet transform
and the coherence between the near-field pressure and far-field
pressure.

\section{Methodology}
\subsection{LES computational setup}
The numerical method was developed and described by
\cite{Kang_2022}. For completeness, the method is briefly
described in this section.

\begin{figure}
    \centering
    \subfigure[\hspace{10cm}\label{a}]{\includegraphics[width=1.9in,angle=0]{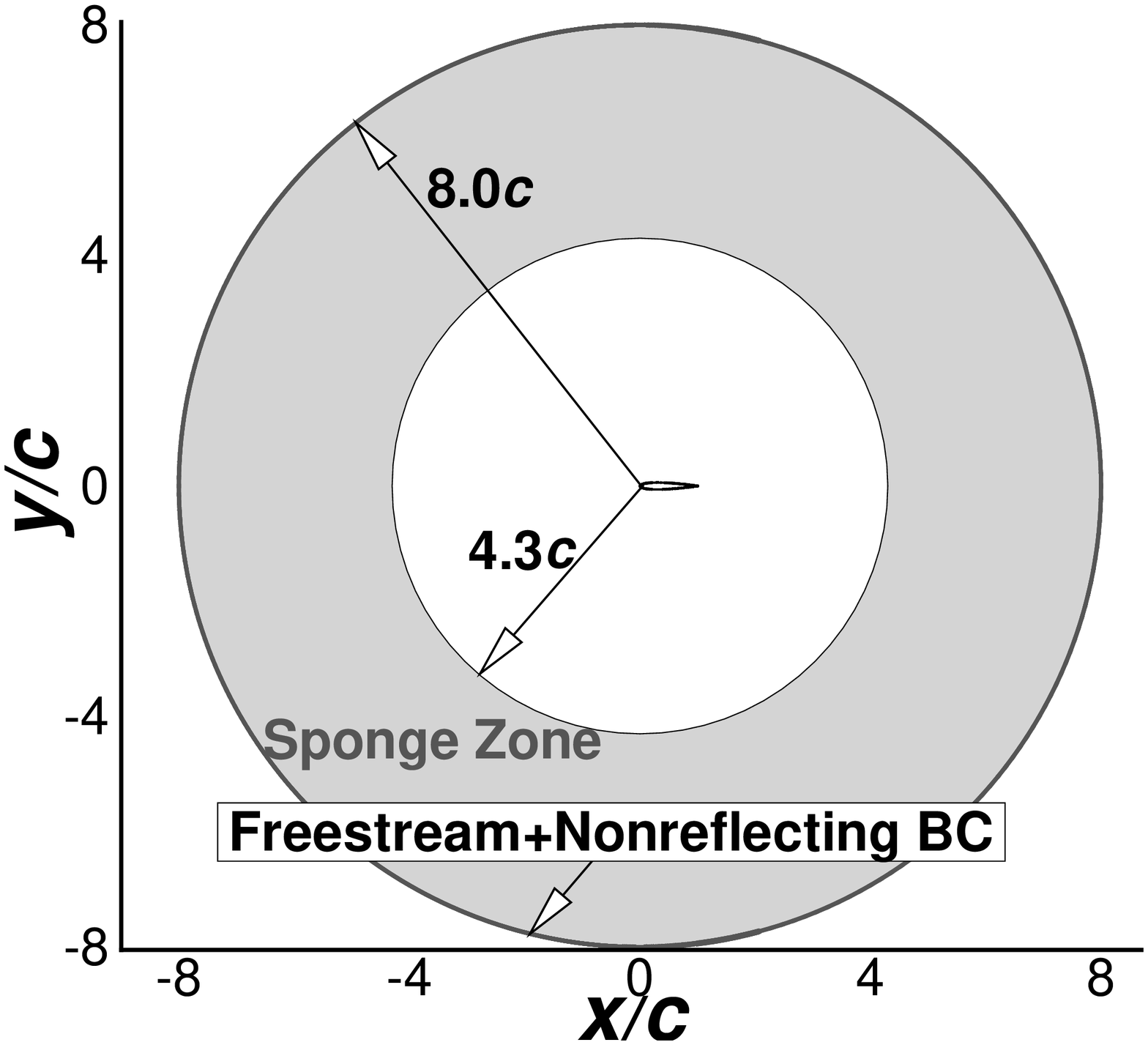}}
    \subfigure[\hspace{10cm}\label{b}]{\includegraphics[width=1.9in,angle=0]{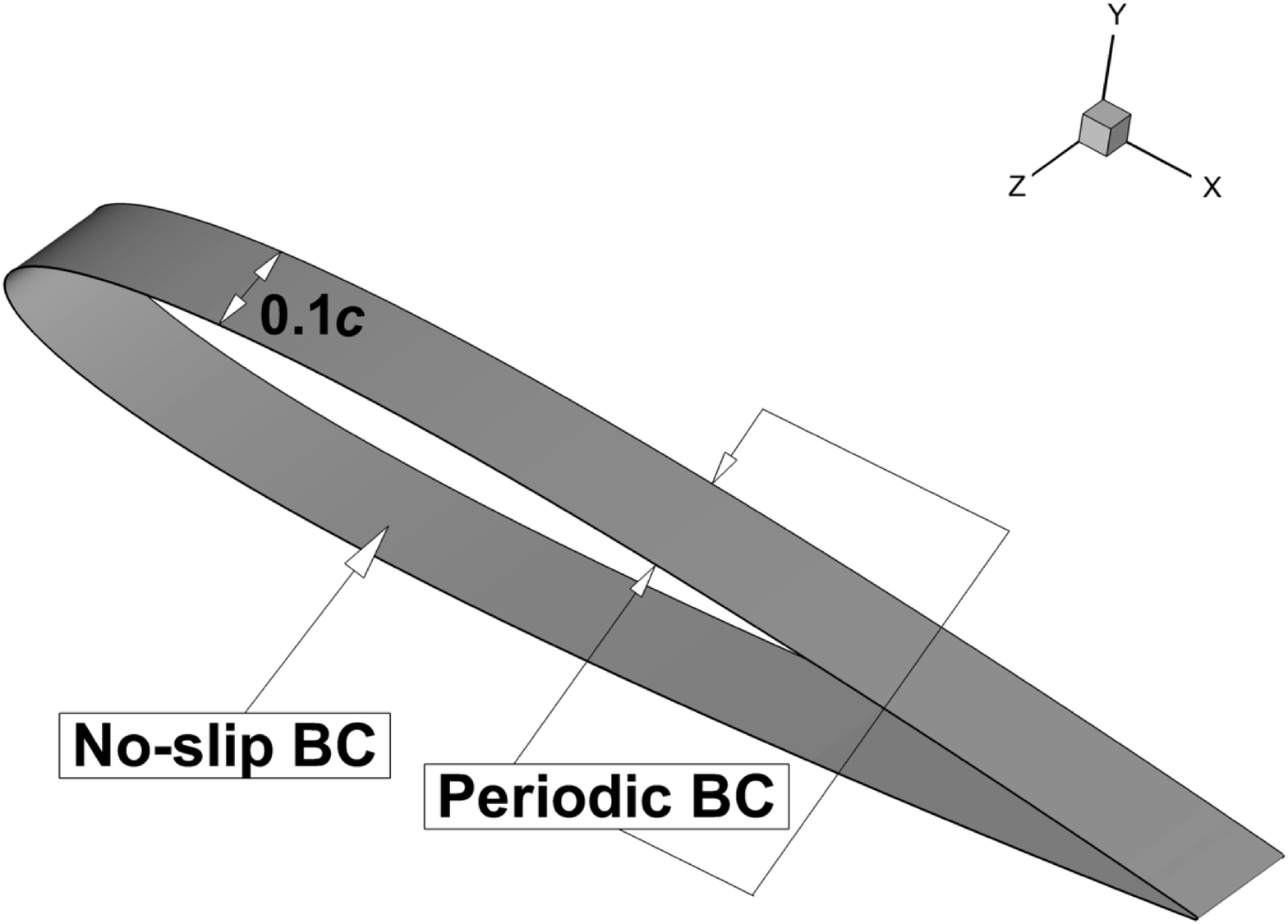}}
    \caption{CFD domain: (a) O-type computational domain over the far field and (b) the three-dimensional airfoil surface with boundary conditions (BCs).}
    \label{fig:Fig1}
\end{figure}

The computational domain and a NACA 0012 airfoil with a blunt
trailing-edge configuration adopted in the present study are
presented in Figure~\ref{fig:Fig1}(a,b). Two angles of attack, $0^{\circ}$
and $6.25^{\circ}$, are considered. The airfoil chord length is
denoted by $c$. A two-dimensional O-grid has a radius of 8.0$c$
and is extruded in the spanwise direction whose length is 0.1$c$.
The no-slip boundary is applied on the airfoil surface. For the
far field, the freestream boundary condition is imposed in which
the Riemann invariant is used depending on the flow direction at
the boundary, and the nonreflecting boundary condition is also
applied in order to prevent sound waves from being reflected from
the far-field domain back to the source of sound. The periodic
condition is applied in the spanwise direction for a
three-dimensional simulation. Finally, a sponge zone is placed
between the radius of 4.3$c$ and the far field to dissipate sound
energy and minimize reflections. Wave strength is gradually
reduced in the sponge zone. The damping coefficient in the
artificial damping term introduced in the Navier-Stokes equations
is defined as $\nu_{max}$ whose optimized value is prescribed as 2 MHz.



\begin{figure}
    \centering
    \subfigure[\hspace{10cm}\label{a}]{\includegraphics[width=1.7in,angle=0]{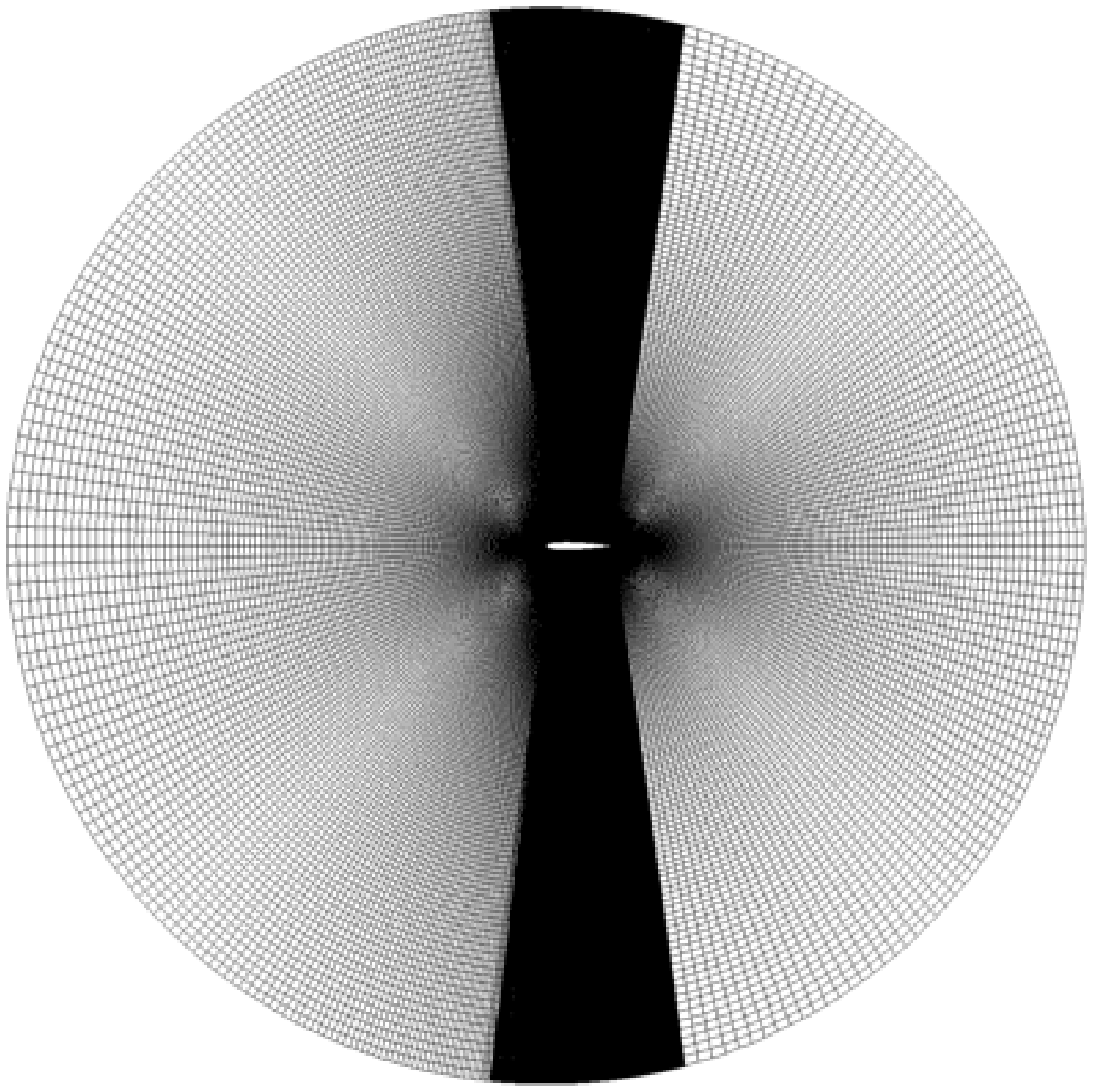}}
    \subfigure[\hspace{10cm}\label{b}]{\includegraphics[width=1.7in,angle=0]{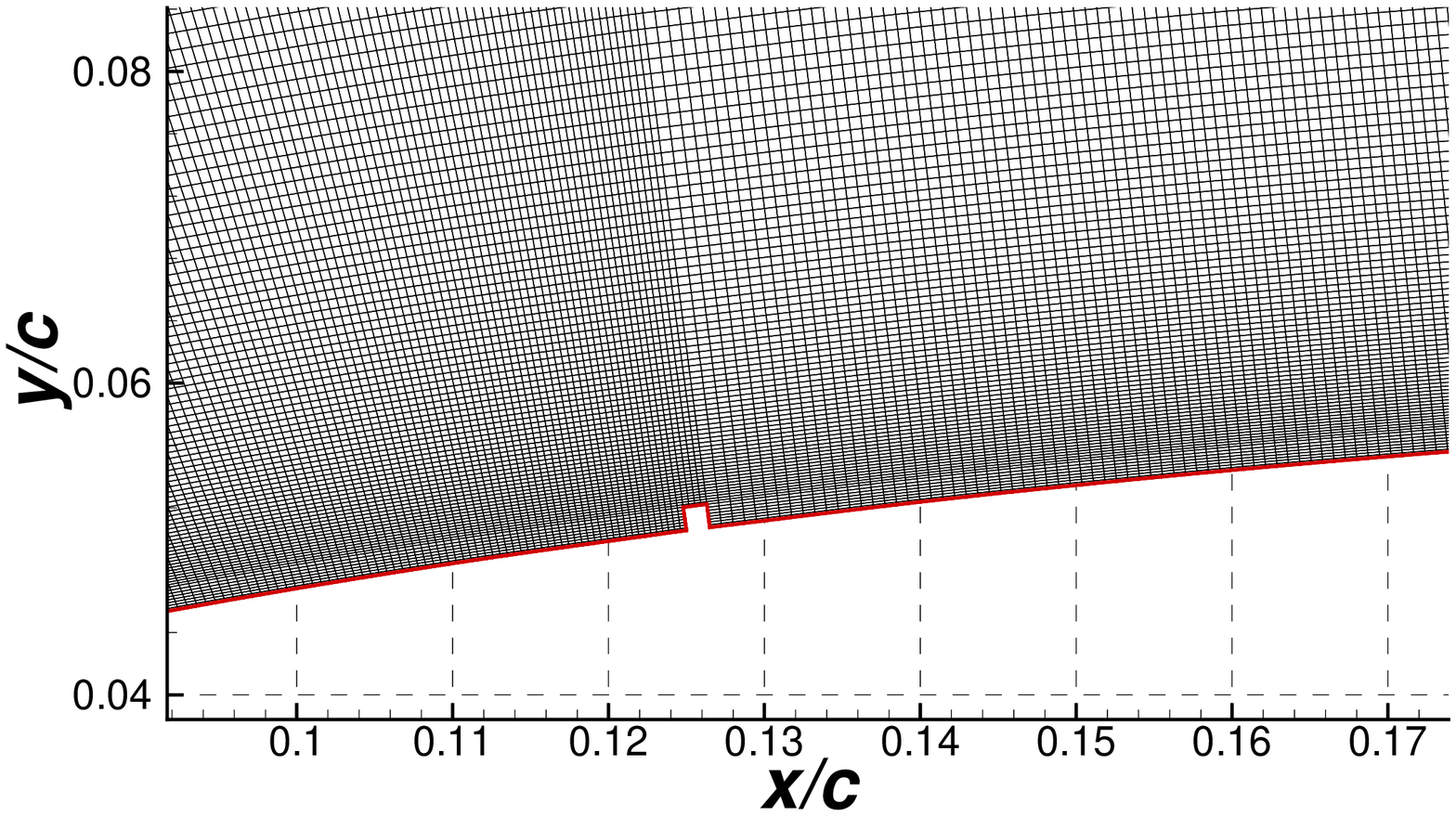}}
    \subfigure[\hspace{10cm}\label{c}]{\includegraphics[width=1.7in,angle=0]{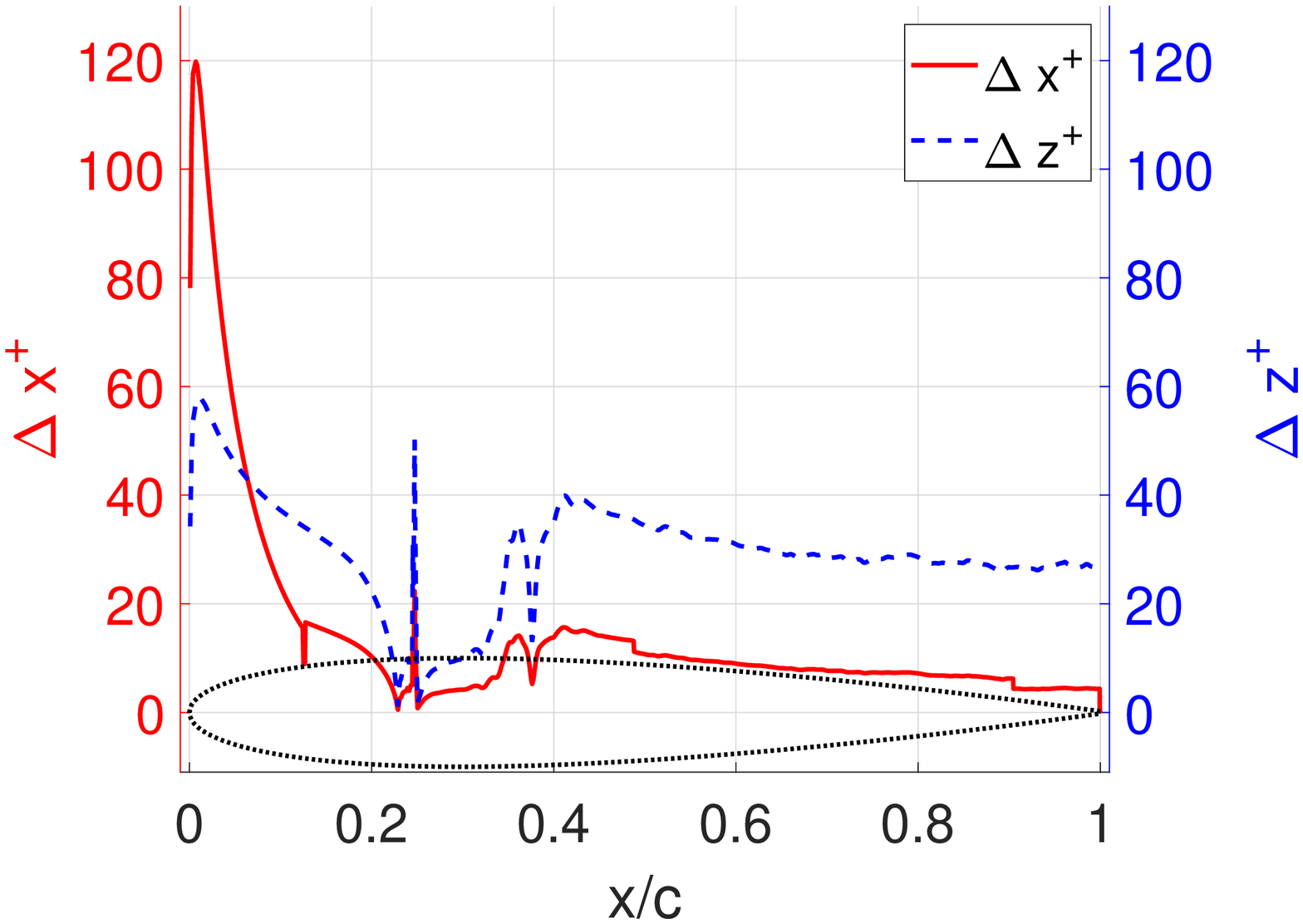}}
    \caption{CFD mesh: (a) structured mesh over the whole domain as a plane view, (b) enlarged near-wall view around the stair-shaped tripdot, and (c) grid spacing in wall units along the suction side of the airfoil.}
    \label{fig:Fig2}
\end{figure}

Using Gmsh \citep{Gmsh_2009}, the structured mesh is generated to span from the
NACA 0012 to the O-grid whole domain, as shown in
Figure~\ref{fig:Fig2}(a,b). The size of the cell distribution is
$N_{x} \times N_{y} \times N_{z} = 4308 \times 323 \times 65$ in
the streamwise, normal, and spanwise directions, respectively, including both
the near-wall and background fields. The grid distribution on the
suction and pressure sides is symmetric; thus, the total number
of grid points is 90,442,300. Two transition cases are considered:
stair-strip enforced transition and natural transition. For the
forced transition, the tripdot is placed at $x/c=0.125$ on both
the suction and pressure sides to trigger the transition to
turbulent flows. One can see the stair-shaped tripdot installed on
the suction side in Figure~\ref{fig:Fig2}(b). The non-dimensional
stair-strip height is 60 viscous wall unit length, which is
between smooth and fully rough cases. The grid is designed to
resolve the near-wall flow fields and thus capture acoustic
sources. Approximately 120 grid points are generated within the
boundary layer near the trailing edge. Figure~\ref{fig:Fig2}(c) shows that $\Delta x^{+}$ is less
than 20 and $\Delta z^{+}$ less than 40 downstream of the tripdot
point. Note that the grid spacing in the wall units is drawn along the
suction side only. Furthermore, $y^{+}$ is maintained at less than
0.1 along the entire airfoil surface in the present simulation. As
shown by \cite{Kang_2022}, the present grid resolution is
satisfactory for wall-resolved LES
\citep{Lee_2021}.

Compressible LES is conducted for the flow past NACA 0012 with a
blunt trailing edge. Finite-volume-based rhoPimpleFoam built in
OpenFOAM (v2012) \citep{OpenFOAM_1998} is employed to simulate unsteady compressible
flows. The spatial discretization is performed using the Gauss
linear scheme, and the temporal discretization uses the
backward-differencing scheme, both of which are second-order accurate. The time step is set to be $10^{-6}$ s, corresponding
to a maximum Courant-Friedrichs-Lewy number of 0.9. The inner loop per one physical
time step is iterated until the convergence criterion is
satisfied, which is $10^{-10}$ in this study. Eddies greater than
the grid size are fully resolved, whereas eddies smaller than the
grid size are modeled by incorporating wall-adapting local eddy
viscosity in the governing equation as a closure model. The
initial condition for LES is obtained from the steady-state
Reynolds-averaged Navier-Stokes simulation for which $k$--$\omega$ shear stress transport is adopted as a closure model. The steady-state
flow solver used is rhoSimpleFoam. The total simulation time in
LES corresponds to 20 airfoil flow-through times (FTT). The data
for spectral processing and acoustic computation are collected
from the last 10 FTT in which flows reach a statistically
convergent state. The parallel computations are made over
computational domains decomposed through OpenMPI, and the elapsed
calculation time was roughly 148,043 s per FTT using 256 cores on
CPU 3.3 GHz, which corresponds to 210,550 total CPU-hours, at the
UC Davis High-Performance Computing Core Facility.

\subsection{Wavelet-based decomposition}
The continuous wavelet transform of a random pressure
fluctuation in the time domain $p(t)$ consists of a projection
over a basis of compact support functions obtained by dilation and
translation of the mother wavelet $\Psi(t)$, which is localized in
both the time and frequency domains. The resulting wavelet
coefficient is a function of time $t$ and of the scale $s$, the
latter being inversely proportional to frequency. The continuous
wavelet coefficient is defined as:
\begin{equation}
    \begin{aligned}
    w(s,t)=C_{\Psi}^{-1/2}s^{-1/2} \int_{-\infty}^{\infty} \Psi^{*}\left( \frac{\tau-t}{s} \right) p(\tau) \,d\tau,
    \end{aligned}
\end{equation}
where $\Psi^{*}((t-\tau)/s)$ is the complex conjugate of the
dilated and translated $\Psi(t)$ and $C_{\Psi}^{-1/2}$ is obtained
satisfying the admissibility condition
\begin{equation}
    \begin{aligned}
    C_{\Psi}=\int_{-\infty}^{\infty} |\omega|^{-1} |\hat{\Psi}(\omega)|^{2}  \,d\omega <\infty.
    \end{aligned}
\end{equation}
Here $\hat{\Psi}(\omega)$ is the Fourier transform of $\Psi(t)$,
\begin{equation}
    \begin{aligned}
    \hat{\Psi}(\omega)=\int_{-\infty}^{\infty} \Psi(t) e^{-j\omega t}  \,dt.
    \end{aligned}
\end{equation}

For a discretized domain as in the present study, a discrete
wavelet transform (DWT) is utilized in practice. A stationary time
series of pressure fluctuation is decomposed by DWT. The discrete
wavelet coefficients are given by
\begin{equation}
    \begin{aligned}
    w_{p}^{(s)}(n)=\sum_{k=-\infty}^{+\infty}{\Psi^{s}(n-2^{s}k)p(k)},
    \end{aligned}
\end{equation}
where $s$ denotes the discretized scale, while the wavelet
function ${\Psi}^{s}(n-2^{s}k)$ is the discretized version of
$\Psi^{s}=2^{-s/2} \Psi\left( t/2^{s} \right)$.
The wavelet kernel used is the Daubechies-12 type. The wavelet
coefficients obtained from DWT are used in a separation algorithm
to isolate the coherent part of the pressure from the incoherent
part of the pressure. A nonlinear recursive denoising technique
is used in the present work \citep{Azzalini_2005}. The incoherent
source of the pressure is iteratively filtered until the
convergence criterion is satisfied. Originally, based on
statistical reasoning \citep{Donoho_1994}, a threshold was guessed as
\begin{equation}
    \begin{aligned}
    T_{o}=\sqrt{2\langle p^{'2} \rangle \log_{2}N_{s}},
    \end{aligned}
\end{equation}
where $\langle p^{'2} \rangle$ is the variance of the pressure
signal and $N_s$ is the number of samples. Starting from the
initial guess above, the threshold is updated at each iteration in the loop whose formulation can be written as
\begin{equation}
    \begin{aligned}
    T_{i}=\sqrt{2\langle p_{I}^{'2} \rangle\vert_{i} \log_{2}N_{s}},
    \end{aligned}
\end{equation}
where $\langle p_{I}^{'2} \rangle \vert_{i} $ indicates the
variance of the incoherent part of the pressure signal in time at the $i$th iteration. The threshold is converged within 20 iterations
in the present work. Such wavelet analysis has been carried out in
MATLAB.

\section{Results}

\subsection{Flow and acoustic results}
\begin{figure}
 \centering
 \subfigure[\hspace{10cm}\label{a}]{\includegraphics[width=2in,angle=0]{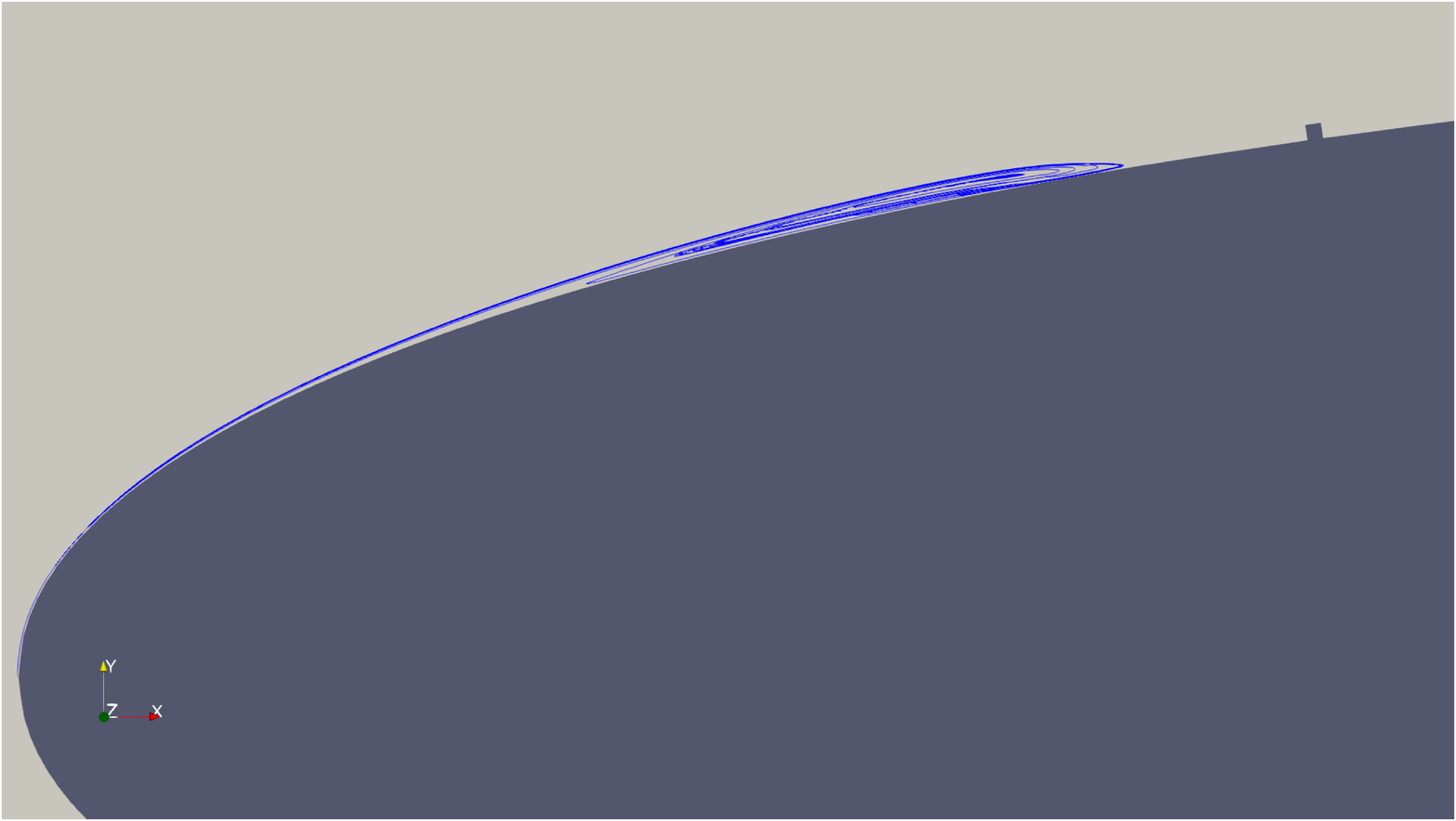}}
 \subfigure[\hspace{10cm}\label{b}]{\includegraphics[width=2in,angle=0]{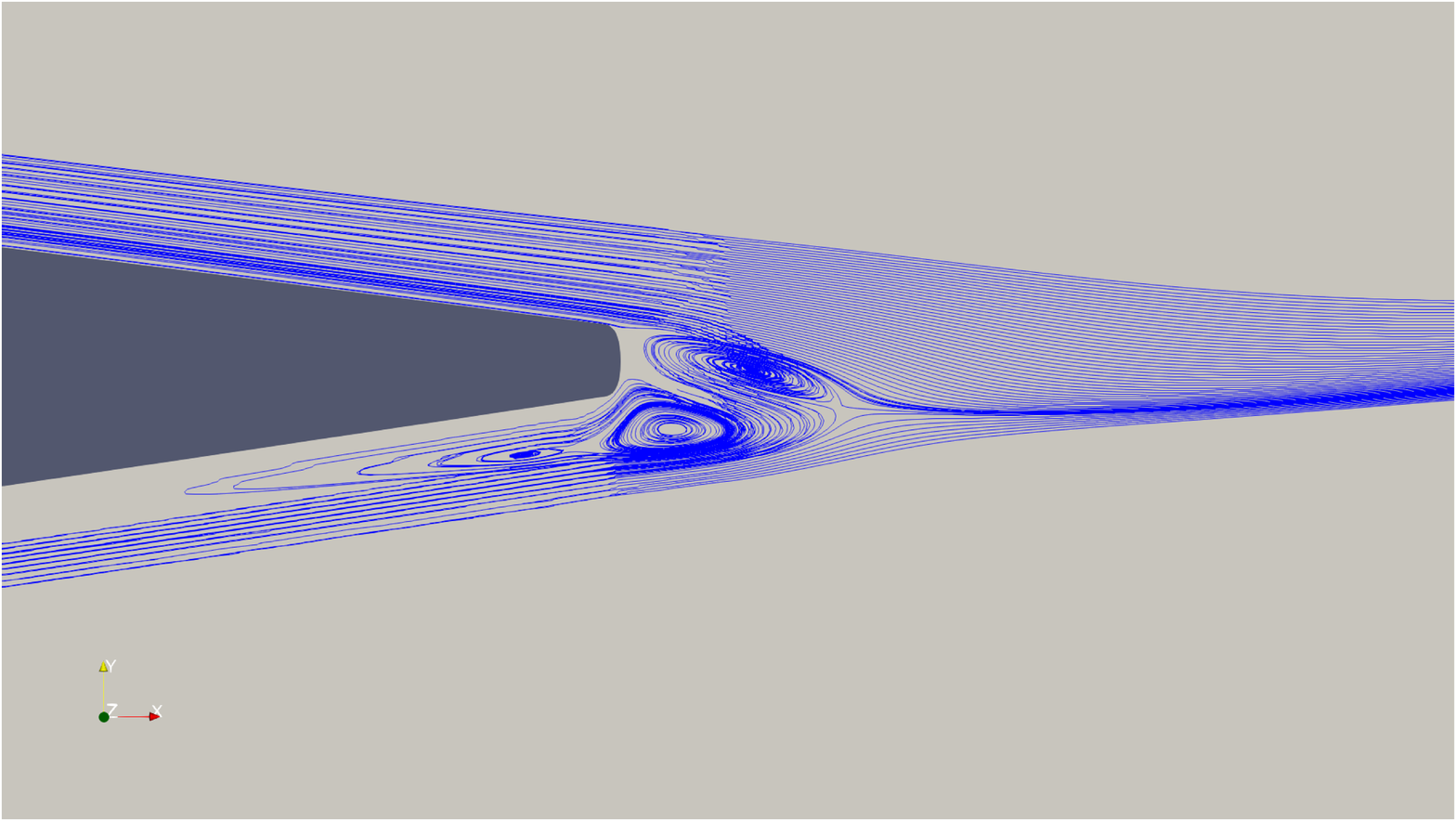}}
 \caption{Streamlines for (a) a laminar separation bubble on the suction side upstream of the stair strip and (b) vortex shedding on the pressure side near the trailing edge.}
 \label{fig:LSBVS}
\end{figure}

The validations of the mean flow and wall pressure spectra at a
zero angle of attack were presented by \cite{Kang_2022}. For the
sake of brevity, the validations are not presented in this paper.
At a $6.25^\circ$ angle of attack, the LSB upstream of the stair strip
on the suction side and vortex shedding near the trailing edge on
the pressure side are observed, as shown in
Figure~\ref{fig:LSBVS}(a,b).

\begin{figure}
 \centering
 {\includegraphics[width=2in,angle=0]{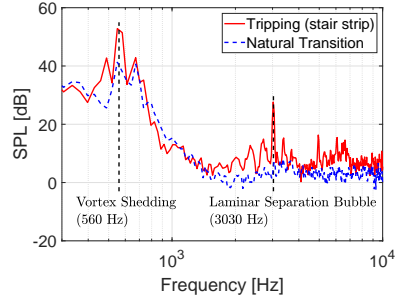}}
 \caption{Sound pressure level (SPL) for tripping and natural transition cases at an angle of attack of $6.25^\circ$.}
 \label{fig:SPL}
\end{figure}

Figure~\ref{fig:SPL} shows the sound pressure level for
stair-strip tripping and natural transition cases at a high angle
of attack. The observer location is
$(x/c,y/c,z/c)=(1.0,0.05,4.07)$ with a chord length, $c$, of 0.3
m. The coordinates $x,y,$ and $z$ denote the streamwise, spanwise,
and wall-normal directions, respectively. Tonal noise appears at
around 560 Hz due to the vortex shedding and at around 3 kHz due
to LSB instability. Although it is not shown in this report, the root mean square of the pressure fluctuations
exhibits a strong peak near the LSB region, which confirms the
noise source location. Although the magnitude is low compared with the tonal noise, the trailing-edge broadband noise occurs beyond 2
kHZ. The visualization of the noise propagation at specific
frequencies will confirm all these noise sources with the aid of
the wavelet-based decomposition method in the next subsection.

\subsection{Wavelet-based decomposed pressures}
\begin{figure}
 \centering
 \subfigure[\hspace{10cm}\label{a}]{\includegraphics[width=2in,angle=0]{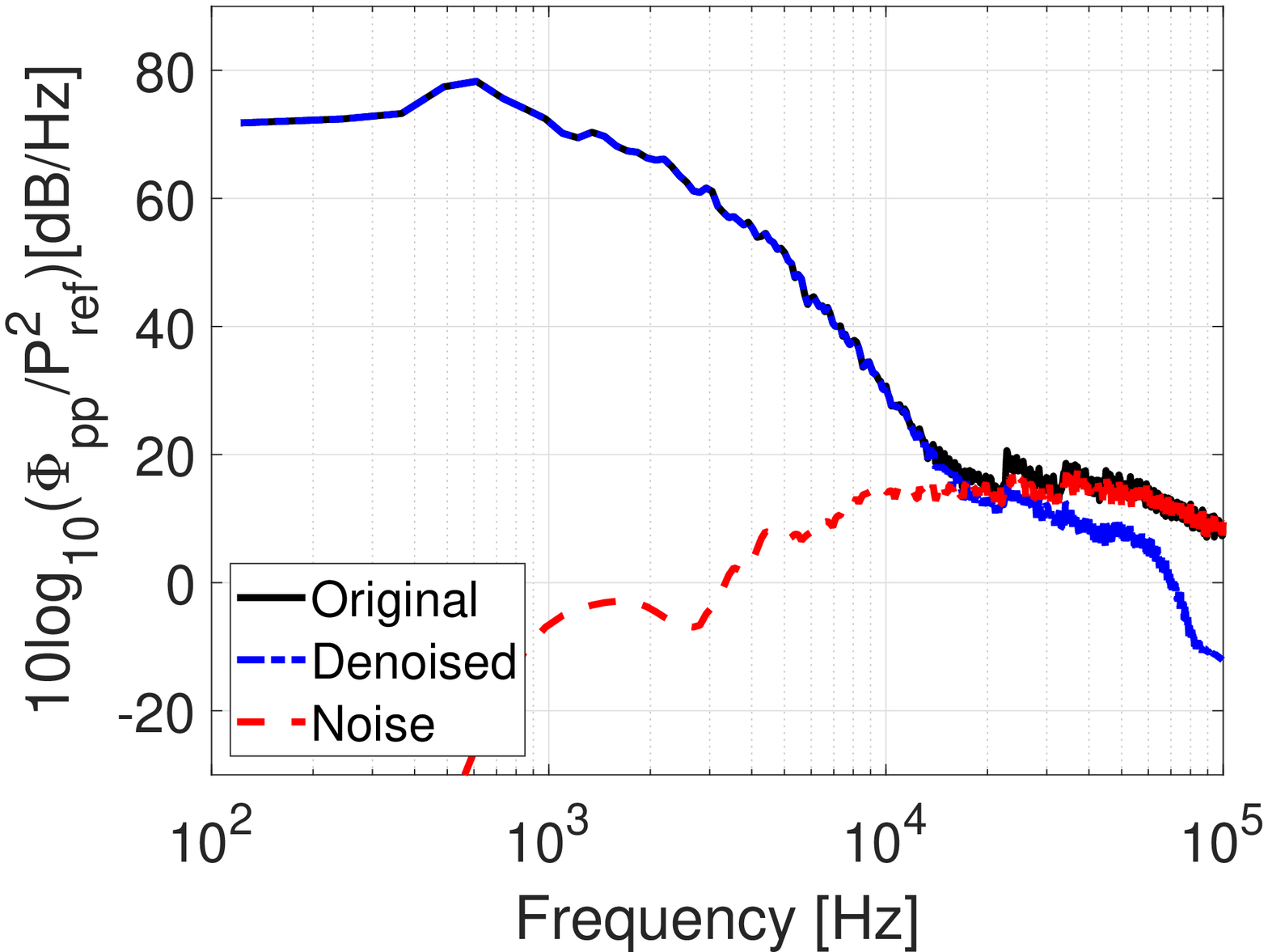}}
 \subfigure[\hspace{10cm}\label{b}]{\includegraphics[width=2in,angle=0]{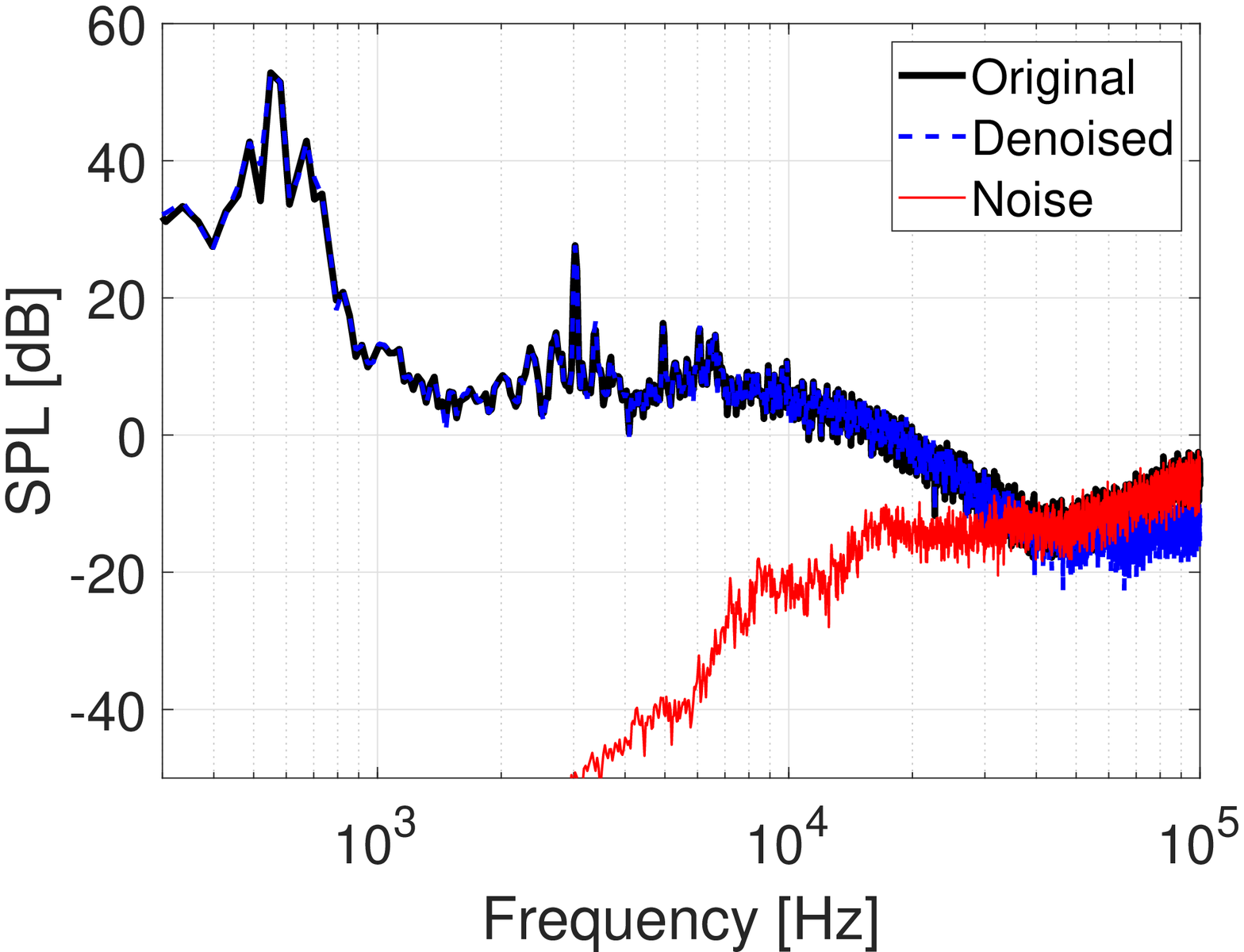}}
 \caption{(a) Decomposed wall pressure spectrum at $x/c=0.99$ and (b) decomposed power spectral density at $(x/c,y/c,z/c)=(1.0,0.05,4.07)$.}
 \label{fig:Decomposed}
\end{figure}

First, the wavelet thresholding technique is used to identify the
numerical noise. Figure~\ref{fig:Decomposed}(a,b) shows the
decomposed wall pressure spectrum at $x/c=0.99$ and the sound power
spectral density at the same observer as that used in
Figure~\ref{fig:SPL}. The Nyquist frequency or half of the
sampling frequency is 500 kHz in both panels. A Hanning window
was used to generate these spectra. The figure shows that the original
wall pressure spectrum has a plateau beyond 20 kHz and is mostly
spurious numerical noise. The magnitude of numerical noise is low
at low frequencies or below 10 kHz, where the denoised or physical
pressure is dominant. This indicates that the numerical noise is
not perfectly white noise. In the SPL plot, the numerical noise contributes only beyond 40 kHz.

\begin{figure}
 \centering
 \subfigure[\hspace{10cm}\label{a}]{\includegraphics[width=2.0in,angle=0]{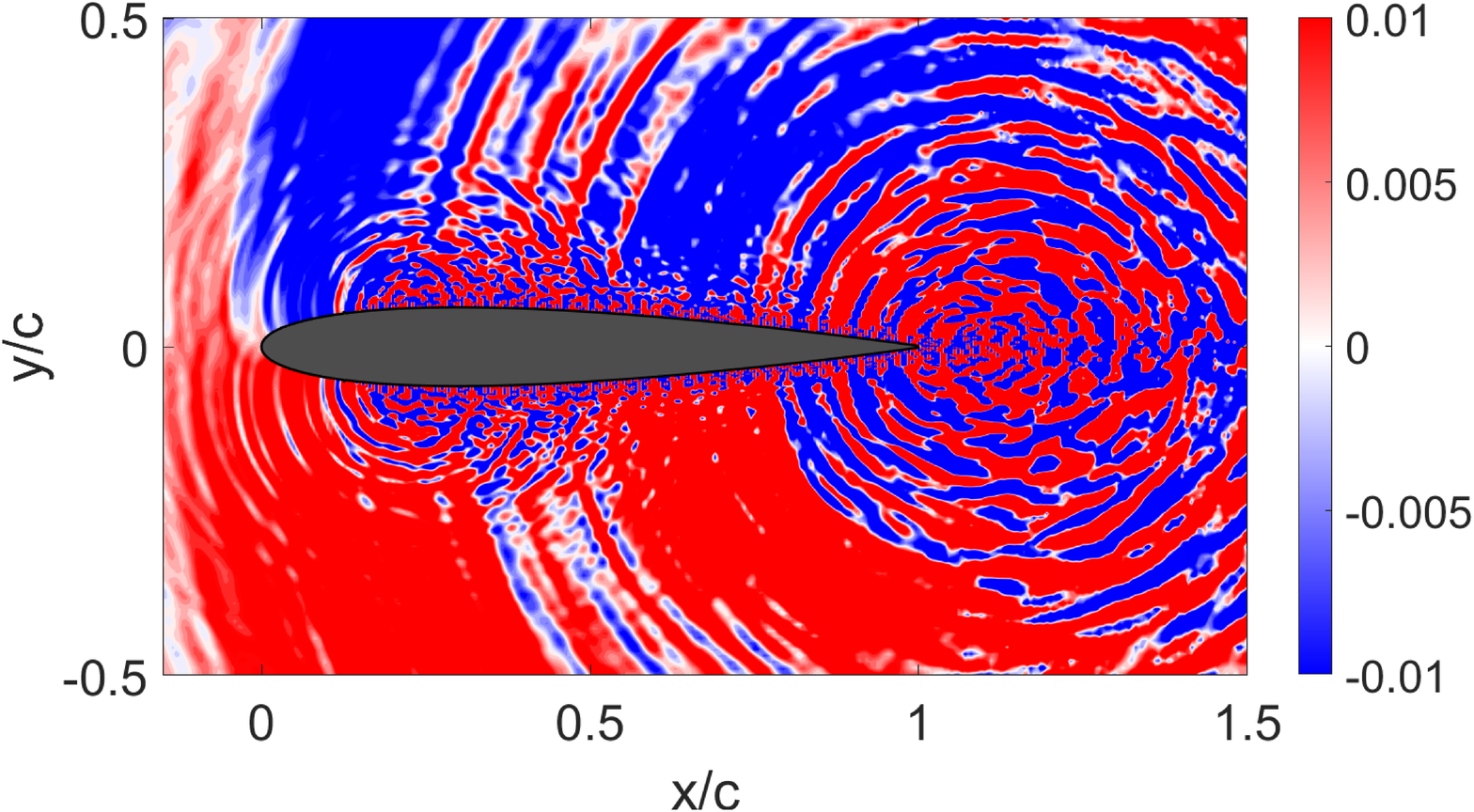}}
 \subfigure[\hspace{10cm}\label{b}]{\includegraphics[width=2.0in,angle=0]{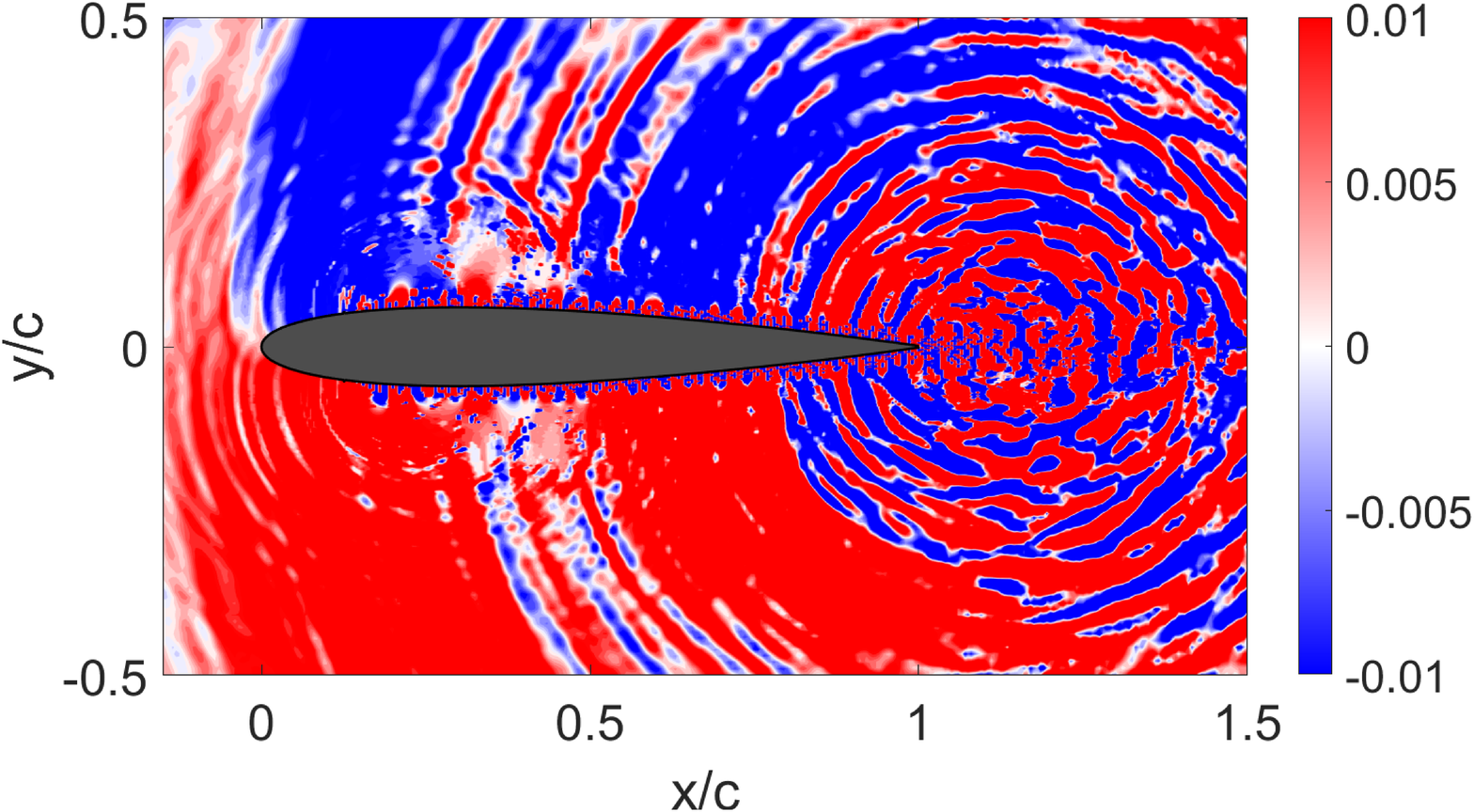}}
 \subfigure[\hspace{10cm}\label{c}]{\includegraphics[width=2.0in,angle=0]{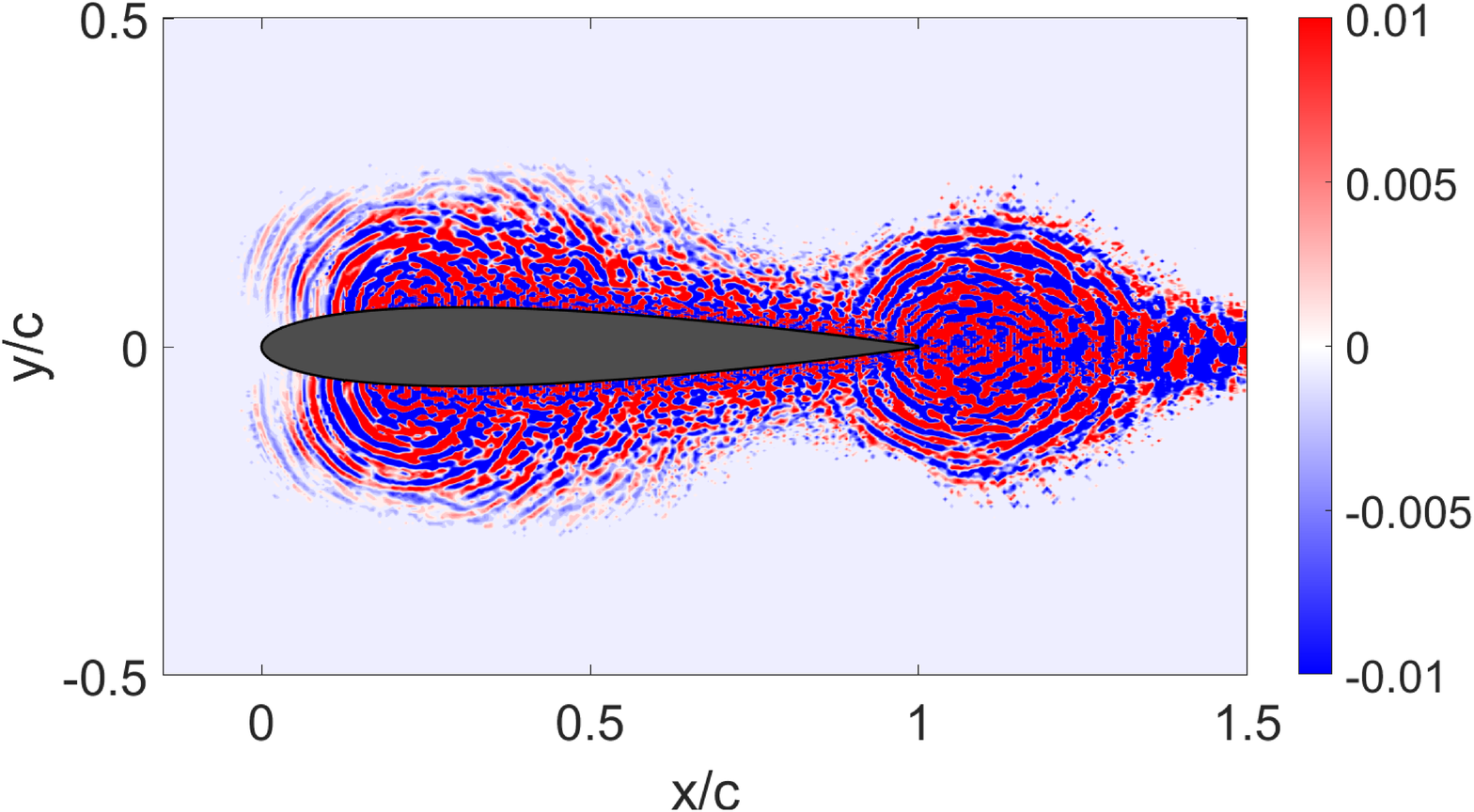}}
 \caption{Dilatation fields cut at midspan for (a) the original pressure, (b) the denoised pressure, and (c) the noise pressure at a zero angle of attack.}
 \label{fig:Dilatation_Decomposed}
\end{figure}

Figure~\ref{fig:Dilatation_Decomposed}(a-c) presents the
dilatation fields of the original pressure as well as the denoised
pressure and the numerical noise pressure at a zero angle of
attack. The numerical noise is evident near the tripping region
and the trailing-edge region. The numerical noise
propagates with the speed of sound so that it can be misunderstood
as the acoustic pressure \citep{Kang_2022}. After eliminating the
numerical noise, the denoised pressure field shows a clearer
picture of the convecting coherent turbulent structures as well as
physical noise propagation. The figure shows that the noise originating
from the tripping region is actually numerical noise so that the
true dominant noise source is trailing-edge noise in this case. In
sum, the wavelet-based decomposition method reveals true noise
source generation and propagation without numerical noise
contamination. Although not shown here, $6.25^\circ$ angle of attack cases show that the numerical noise is not significant
compared with the physical noise so that the original pressure and
denoised pressure look similar. The LSB plays a role in generating
significant physical noise, as shown in Figure~\ref{fig:SPL}, which
is much stronger than the numerical noise.

\begin{figure}
 \centering
 \subfigure[\hspace{10cm}\label{a}]{\includegraphics[width=2in,angle=0]{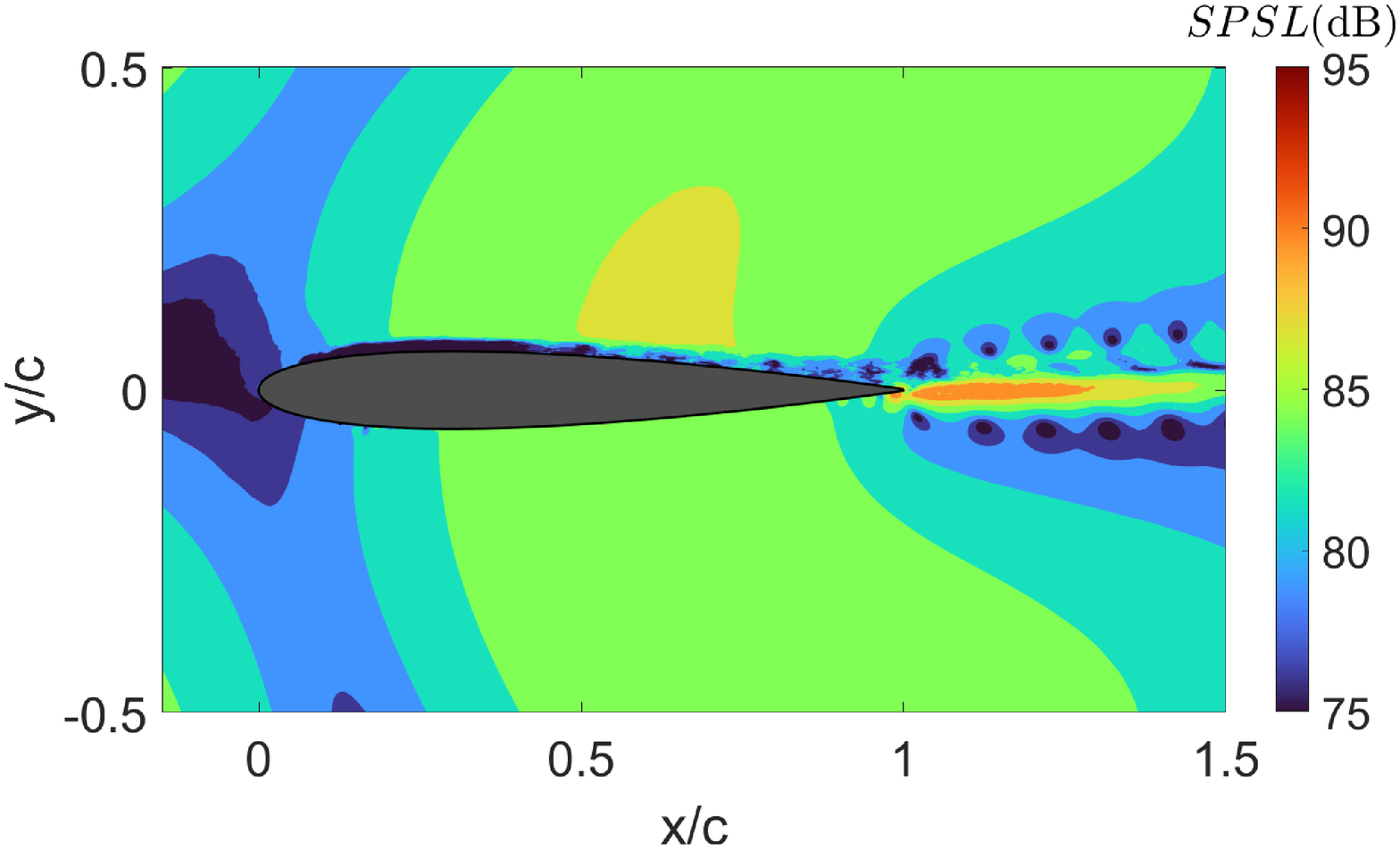}}
 \subfigure[\hspace{10cm}\label{b}]{\includegraphics[width=2in,angle=0]{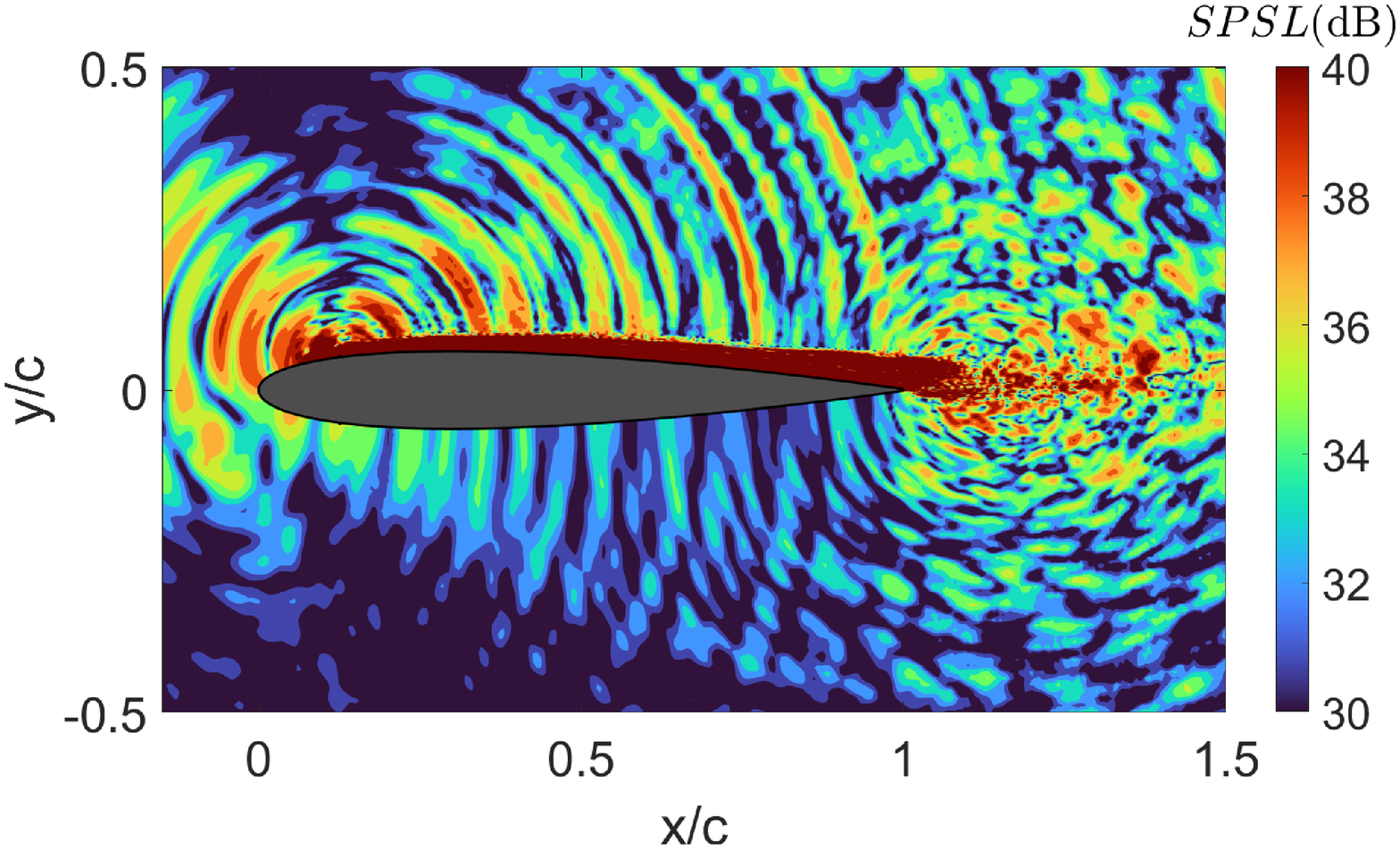}}
 \caption{Contours of the cross-spectrum level using the wavelet-based decomposed pressures with down-sampling: (a) $P_1$ at 500 Hz and (b) $P_2$ at 6 kHz. The reference point for the cross spectrum is located at the trailing edge.}
 \label{fig:SPSL}
\end{figure}

Second, the original pressure is down-sampled with the Nyquist
frequency of 10 kHz before the wavelet transform is applied. The
same thresholding technique is used, where the pressure
reconstructed with the wavelet coefficients whose magnitudes are
larger than the threshold value is denoted $P_1$ and the
remaining pressure is denoted $P_2$. Figure~\ref{fig:SPSL}(a,b)
shows the contours of the cross-spectrum level, defined as $\mathrm{SPSL(dB)}
=10\log_{10} (|S_{xy}| \triangle f/p_{ref}^2)$, between a
point near the trailing edge and points in the two-dimensional field for $P_1$
and $P_2$ at 500 Hz and 6 kHz, respectively, where $S_{xy}$ is the
two-point cross spectrum, $\triangle f=1/T=6.67$ Hz, and
$p_{ref}=20~\mu$Pa. Although not shown here, the cross-spectrum contours of $P_1$ become zero at high
frequencies, while the cross-spectrum contours of $P_2$ become zero
at low frequencies. This indicates that $P_1$ represents the
low-frequency vortex shedding noise and $P_2$ represents the
high-frequency noise that contains the LSB instability noise and
trailing-edge noise. The leading-edge noise
generated from the LSB instability is more evident than the
trailing-edge noise in Figure~\ref{fig:SPSL}(b), even though the
reference point is located at the trailing edge when computing the
cross-spectrum-level contour plots. This indicates that the
pressure at the trailing edge is strongly influenced by the
leading-edge noise or that the leading-edge noise is stronger than the
trailing-edge noise in this case. In sum, the wavelet-based
decomposition technique with down-sampling, that is, without any
high-frequency numerical noise, is useful to separate the noise
sources as well as their acoustic propagation depending on flow
structures or noise sources.

\begin{figure}
 \centering
 \subfigure[\hspace{10cm}\label{a}]{\includegraphics[width=2in,angle=0]{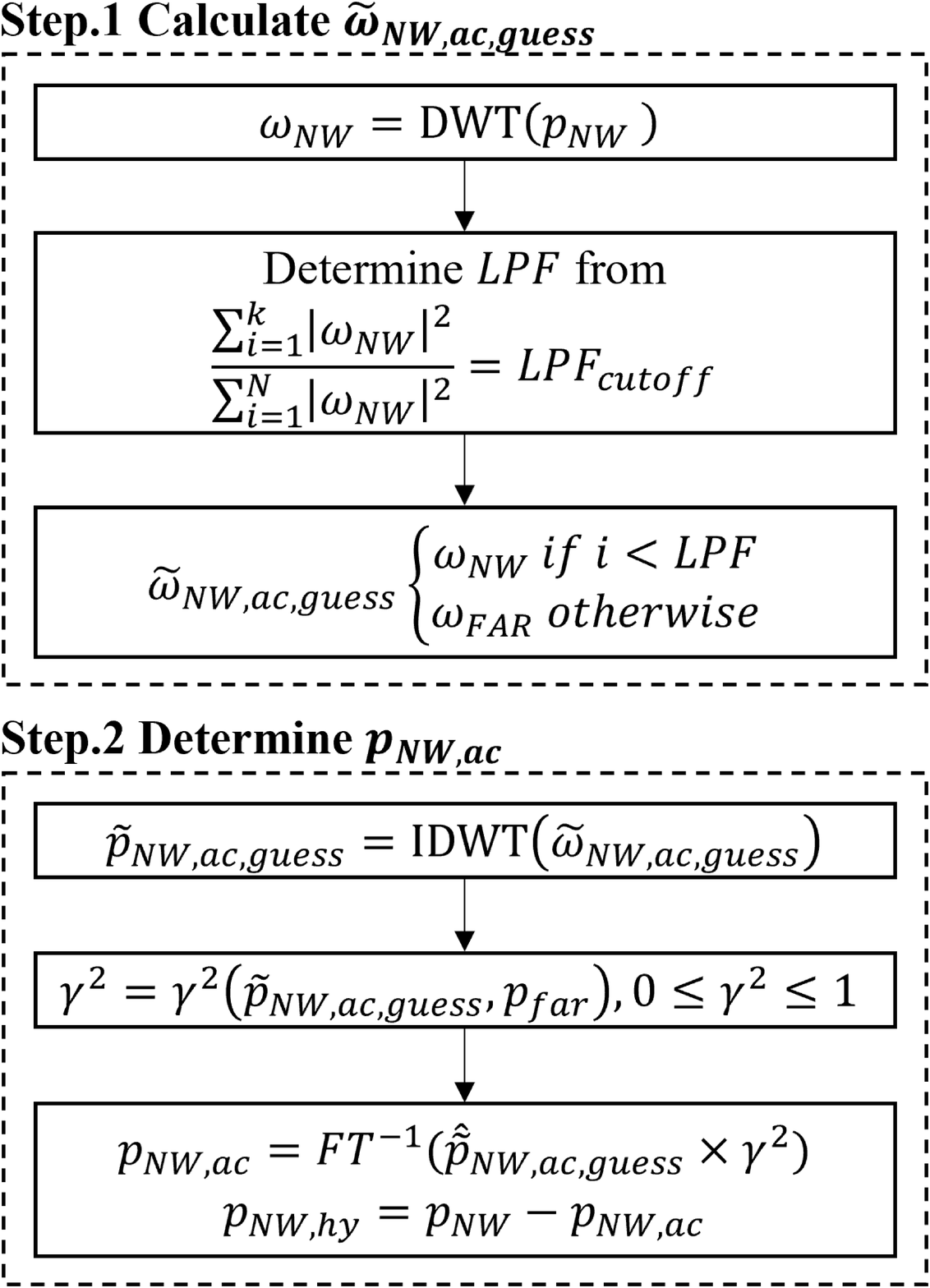}}
 \subfigure[\hspace{10cm}\label{b}]{\includegraphics[width=2.7in,angle=0]{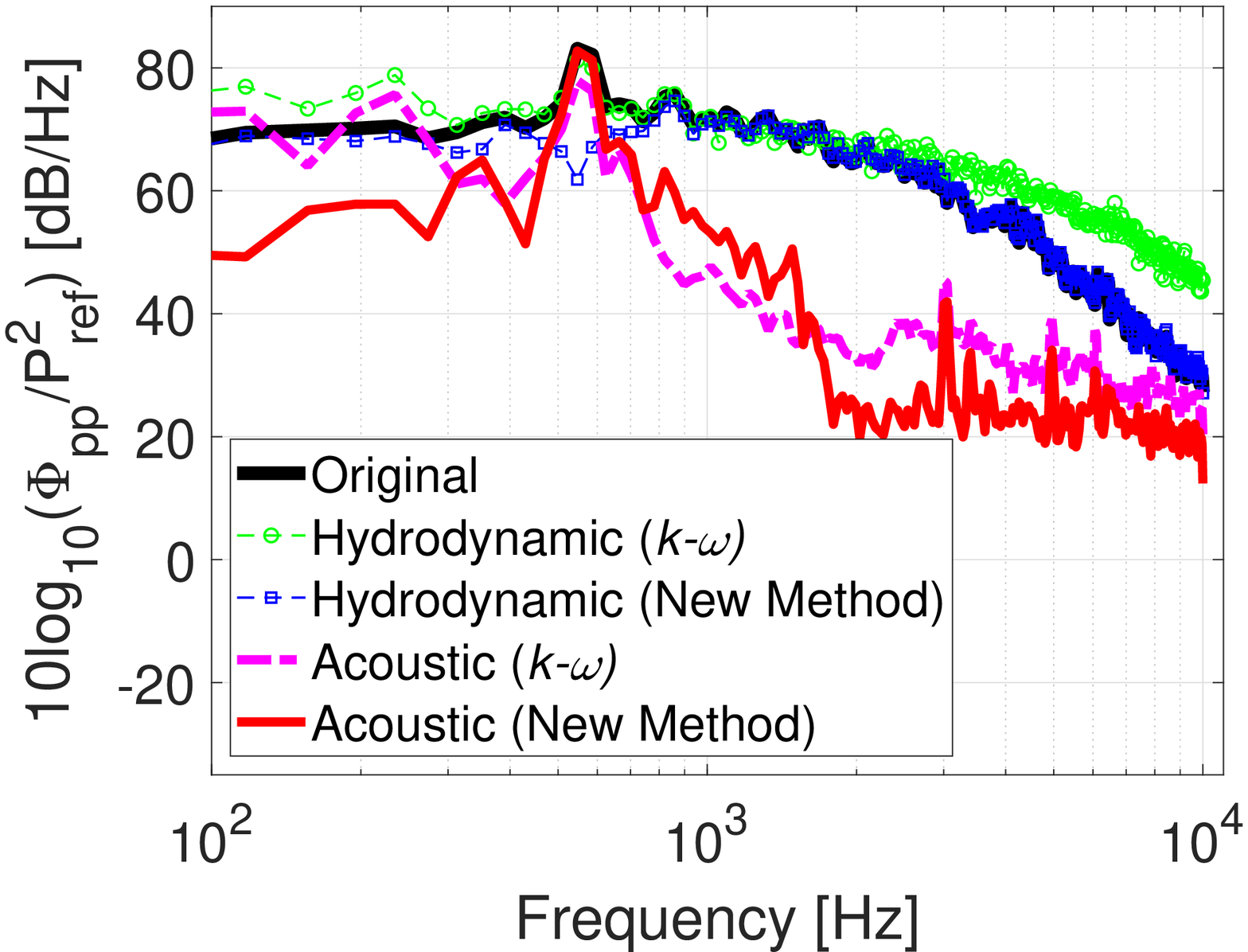}}
 \caption{(a) Flow chart of the numerical algorithm for the new decomposition method and (b) the results of the decomposed pressures near the trailing edge.}
 \label{fig:Phy_Pac_decomposition}
\end{figure}

Finally, the wavelet transform is used to separate the
hydrodynamic pressure from the acoustic pressure near the surface.
The main idea is to use the cross-correlation or coherence between
the pressure near the surface and far-field pressure. A key
assumption is that the acoustic pressure near the wall is strongly correlated with the far-field pressure while the
hydrodynamic pressure is less correlated with the far-field
pressure. There is still a challenge in this approach. The
magnitudes of specific wavelet coefficients from the pressure near
the surface contain both the hydrodynamic and acoustic
contributions so that separation cannot be achieved by just
sorting out the wavelet coefficient magnitudes using the threshold
denoising technique, which was used in jet noise
\citep{Mancinelli_2017}. In our new algorithm, the extent of the
acoustic contribution in the wavelet coefficient magnitude is
determined from the coherence between the pressure near the
surface and far-field pressure.
Figure~\ref{fig:Phy_Pac_decomposition}(a) shows the flow chart of
the algorithm. After performing DWT of the original near-wall
pressure, the index, $k$, of the wavelet coefficient is determined on the basis of the low-pass filter (LPF) cutoff value, which is assumed
to be 0.96 in the current study. This value was
selected when the sum of the wavelet coefficients squared reaches
plateau. Therefore, this cutoff value can be case dependent. The
near-wall wavelet coefficients below this index or the point of
LPF are retained. Above this range, the wavelet coefficients of
the far-field acoustic pressure ($\omega_{FAR}$) are taken due to the dominance of the hydrodynamic pressure in this range for the near-wall pressure. In the next step, the wavelet
coefficients for the guessed near-wall acoustics ($\tilde{\omega}_{NW,ac,guess}$) are used to find
the acoustic pressure through inverse DWT (IDWT). The
coherence between the guessed near-wall acoustic pressure and the
far-field pressure is used to determine the final near-wall
acoustic pressure. Finally, the hydrodynamic pressure
($p_{NW,hy}$) is obtained by subtracting the acoustic pressure
($p_{NW,ac}$) from the total pressure.
Figure~\ref{fig:Phy_Pac_decomposition}(b) shows the original
pressure, hydrodynamic pressure, and acoustic pressure near the
trailing edge on the suction side using this algorithm. For the
validation, the hydrodynamic and acoustic pressures are also
obtained from the wavenumber-frequency-domain Fourier transform
approach. Although the results are not perfectly matched, the new
method shows the correct tonal peaks at 500 Hz, 3 kHz, and 5 kHz.
The acoustic pressure also captures the right trend at high
frequency, which exhibits a rapid decay. The hydrodynamic pressure
in the new method follows the original pressure in almost the
entire frequency range, while the wavenumber-frequency-domain
Fourier transform result shows a higher hydrodynamic pressure than
the original pressure at high frequency, which might be possible
with the destructive interference between the hydrodynamic and
acoustic pressures. In sum, the new method provides promising
results for decomposing the hydrodynamic and acoustic pressures.
Yet, more work is needed to improve the results.


\section{Conclusions}
This paper has applied the wavelet transform approach to airfoil
aerodynamic noise to decompose the pressure in various useful
ways. First, the wavelet-based thresholding denoising technique
identifies and eliminates the nearly Gaussian numerical noise. The numerical noise is dominant near the tripping- and trailing-edge regions at high frequencies. Second, the same
wavelet technique with down-sampling decomposes the acoustic
pressure into low-frequency vortex shedding noise and
high-frequency LSB instability noise as well as trailing-edge
noise. Third, the coherence between near-field pressure and
far-field pressure was applied to the wavelet coefficients to
decompose the hydrodynamic and acoustic pressures in the near
field. Yet, this method needs further improvement to achieve more
precise separation between the two pressure components.

It is expected that this novel technique will provide new
insights into airfoil noise generation and propagation mechanisms.
For example, the separation between the hydrodynamic and acoustic
pressures helps improve and guide the low-fidelity acoustic
prediction model developments. In particular, complex flow and
acoustic interactions with passive and active noise control means
can be clearly revealed with this technique.

\subsection*{Acknowledgments}
The authors acknowledge use of computational resources from the
Yellowstone cluster awarded by the National Science Foundation to
CTR. The authors appreciate valuable discussions with Prof. Kai
Schneider about the wavelet technique during the CTR program.
\bibliography{references}

\end{document}